\documentclass [aps,pra,amssymb,amsmath,twocolumn,showpacs]{revtex4-1}

\usepackage{amsmath}
\usepackage{amssymb}
\usepackage{graphicx}
\usepackage{verbatim}
\usepackage{bm}
\usepackage{color}

\usepackage{graphicx,epsfig,amsfonts,amssymb}
\usepackage{bm}% bold math
\usepackage{times}
\usepackage{lipsum}

\usepackage[all]{xy}

\newcommand{\la}{\langle}
\newcommand{\ra}{\rangle}
\newcommand{\rar}{\rightarrow}

\newcommand{\be}{\begin{eqnarray}}
\newcommand{\ee}{\end{eqnarray}}

\newcommand{\bs}{\begin{equation}\begin{split}}
\newcommand{\es}{\end{split}\end{equation}}

\date{\today}

\begin{document}
\title{Cooperative Light Emission in the Presence of Strong Inhomogeneous Broadening}

%\author{Chen Sun$^{a, b}$, Vladimir Y. Chernyak$^{c, d}$, Nikolai A. Sinitsyn$^a$, and Emil Yuzbashyan$^e$}
\author{Chen Sun$^{a}$, Vladimir Y. Chernyak$^{b,c}$, Andrei Piryatinski$^d$, Nikolai A. Sinitsyn$^d$ }
%\affil[1]{ Theoretical Division, Los Alamos National Laboratory, Los Alamos, NM 87545,  USA}
\affiliation{$^a$Department of Physics, Brown University, Providence, Rhode Island 02912, USA}
\affiliation{$^b$Department of Chemistry, Wayne State University, 5101 Cass Ave, Detroit, Michigan 48202, USA}
\affiliation{$^c$Department of Mathematics, Wayne State University, 656 W. Kirby, Detroit, Michigan 48202, USA}
\affiliation{$^d$Theoretical Division, Los Alamos National Laboratory, Los Alamos, NM 87545, USA}

%\affiliation{$^e$Center for Materials Theory and Department of Physics and Astronomy, Rutgers University, Piscataway, NJ 08854, USA}

\begin{abstract}
We study  photon emission by an ensemble of two-level systems, with strong inhomogeneous broadening and coupled to a cavity mode whose frequency has linear time-dependence.  The analysis shows that, regardless the distribution of energy level splittings, a sharp phase transition occurs between the weak and strong cooperative emission phases near a critical photonic frequency sweeping rate. The associated scaling exponent is determined.  We suggest that this phase transition can be observed in an ensemble of negatively charged NV$^-$ centers in diamond interacting with a microwave half-wavelength cavity mode even in the regime of weak coupling and at strong disorder of two-level splittings.
\end{abstract}

\maketitle
For a system driven by time-dependent fields, one can define the limits of fast and slow (adiabatic) field  modulation. In the former limit, the system does not have time to respond. In the latter  limit, the system adjusts to be in an instantaneous eigenstate of the Hamiltonian. The intermediate regime, however, is usually poorly understood; thus it holds potential to identify novel and robust effects allowing for new functionalities of solid-state microstructures.

%However, it  is usually quite difficult to get an insight into the dynamics happening in the driving regime in between these limits. 

In this letter, we examine a quantum many-body model whose behavior depends discontinuously on a parameter describing the rate of time-dependent control. Depending on this parameter, the system behaves  as  driven  either very quickly or almost adiabatically. The transition between these phases is marked by a critical modulation rate such that an order parameter can be defined that is zero in one phase and grows following a power law in the other phase. %Moreover,  the behavior near this phase transition does provide a new functionality to a mesoscopic solid-state device. 
Thus, we demonstrate that the limiting dynamic characteristics such as ``fast" and ``slow" can be attributed to different phases divided by a sharp boundary belonging to the intermediate modulation regime.
%Such a phenomenon is a dynamical analog of the phase transition between the para- and ferromagnetic states describing spin systems in thermodynamic equilibrium. 
 
Specifically, we consider a generalized Tavis-Cummings (TC) Hamiltonian:
\be
\hat H=\omega   \hat{a}^{\dagger}\hat{a} +\sum_{k=1}^N \varepsilon_k \hat{\sigma}_k^z + \sum_{k=1}^N g_k(\hat{a}^{\dagger} \hat{\sigma}_k^- +\hat{a} \hat{\sigma}_k^+),
\label{dtcm1}
\ee
where $\omega$ is the frequency of a bosonic mode $\hat{a}$; $\hat{\bm\sigma}_k$ are the Pauli operators of $k$-th (out of $N$) spin;  $\varepsilon_k$ and $g_k$ are the spin level splitting energy and the coupling constant to the bosonic mode, respectively. This model is notorious for predicting a variety of quantum cooperative effects in an ensemble of two-level quantum emitters coupled to a photonic cavity mode \cite{TC-general}. However, for  solid-state quantum emitters, local strains and dipole fields randomize energy level splitting $\varepsilon_k$ leading to the inhomogeneous broadening. This results in the coherence loss disrupting the cooperative effects  at weak coupling, i.e., when $\sqrt{{\rm var}(\varepsilon_k)} \gg \la g_k \ra \sqrt{N}$ \cite{inhom-super}. 

%%%%%%%%%%%%%%%%%%%%%%%%%%%%%%%%%%%%%%%%%%%%%
\begin{figure}
\scalebox{0.35}[0.35]{\includegraphics{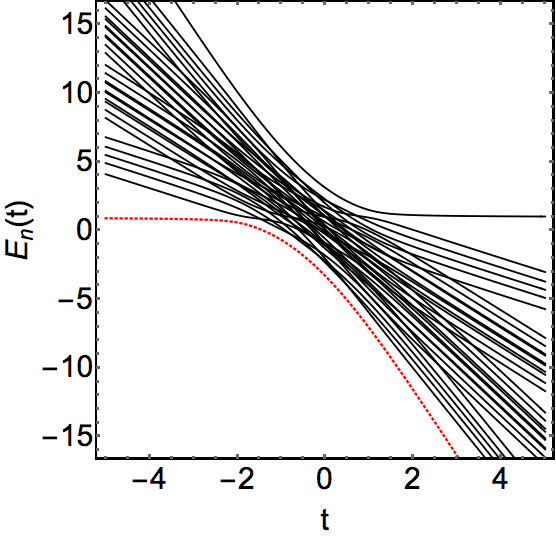}}
\caption{The adiabatic energy levels of the Hamiltonian (\ref{dtcm1}) in the sector with five up-polarized spins ($2^5=32$ states in the phase space) and zero bosons in the initial state; $\omega=-\beta t$. 
The red dotted curve marks the ground energy level for this sector. Here $\beta=1$, $g_k=0.25, \, \forall k$ and $\varepsilon_k=0.8\cdot (k-N/2+r_k)$, where $r_k$ are random numbers from the range of $(0,0.5)$. }
\label{spectrumTC-fig}
\end{figure}
%%%%%%%%%%%%%%%%%%%%%%%%%%%%%%%%%%%%%%%%%%%%%
To address this problem, we set the cavity mode frequency to vary in time linearly around the mean, $\la {\varepsilon}_k\ra$, of the two-level splitting with a rate $\beta>0$:
\be
\omega (t) = \la {\varepsilon}_k\ra -\beta t,
\label{linearT}
\ee
and require that as $t$ increases from a large negative, $-T$, to a large positive, $T$, values,  the photon frequency passes through all spin resonances, i.e., $|\pm\beta T|>\sqrt{{\rm var}(\varepsilon_k)}$.  Note that similar protocols are used for binding ultracold atoms into molecules \cite{bose1,bose2}.

For the slow modulation limit, $\beta \rar 0$, the adiabatic theorem guarantees that the system remains in the ground state if its energy is separated by a gap from the rest of the spectrum. The Hamiltonian (\ref{dtcm1}) conserves the number of excitations represented by the up-spins plus bosons. Thus for $N$ excitations, the initially fully up-polarized spin state is the ground state. In Fig.~\ref{spectrumTC-fig} we show that this state changes so that its energy is always separated by a gap from other levels.  Hence, we can exclude the scaling arguments of the Kibble-Zurek theory \cite{kzt1,kzt2,kzt3}, and  nonadiabatic excitations become suppressed exponentially for sufficiently small $\beta$. After the slow linear sweep of the frequency (\ref{linearT}), all spin excitations are transferred to the photons that become far off-resonance at the end. 
%{\color{red}The light coherence time then is determined mainly by the cavity quality factor. } 

{However, the adiabatic limit is usually hard to achieve. As a practical system, we  consider negatively charged NV$^-$ centers in diamond coupled to a half-wavelength size microwave cavity mode.} Then the basic parameters are $\la {\varepsilon}_{k} \ra \sim 3$GHz, $ \sqrt{{\rm var}(\varepsilon_k)}\approx 3$MHz,  $g_k\sim10$Hz \cite{nv-strong}. {We intentionally do not consider a system with artificially reduced disorder or increased coupling. The initial spin polarization can be induced optically.  For example, there is  a maser based on NV$^-$ centers at room temperature \cite{nv-maser1}.} It was also shown in \cite{nv-lifetime} that if the cavity mode is  far off-resonance at cryogenic temperature, the spin lifetime $T_1$ increases to many hours. Spin coherence time $T_2$ for well isolated NV-centers reaches milliseconds at room temperature \cite{long-coherence-NV1},  and such a quantum lifetime is found  generally at liquid nitrogen temperatures \cite{long-coherence-NV2}. 
Moreover, it is possible to sweep frequency of a cavity within $1$GHz frequency range using the piezoelectric effect \cite{nv-tunable}. Thus, our assumptions about the Hamiltonian (\ref{dtcm1}), the initial conditions, and the protocol (\ref{linearT}) can be experimentally realized.

The condition for the adiabatic limit is $ \la g_k^2 \ra /\beta \gg 1$, which requires  the passage  through the inhomogeneously broadened region during time $2T\approx  2\sqrt{{\rm var}(\varepsilon_k)}/\beta\sim 10^5$sec. This should be compared with the  lifetime of a photon in the cavity $\tau_{c} =Q/\la {\varepsilon}_{k} \ra$. The quality factor $Q$ is usually not exceeding values above $Q\sim 10^6$ \cite{nv-largeQ}, for which $\tau_{c}\approx10^{-4}$sec. At extra cost, it is possible to reach $Q\sim 10^9$ \cite{ultrahighQ} but this is still far from what is needed to reach the adiabatic regime without photon loss.

%%%%%%%%%%%%%%%%%%%%%%%%%%%%%%%%%%%%%%%%%%%%%
\begin{figure}
\scalebox{0.5}[0.5]{\includegraphics{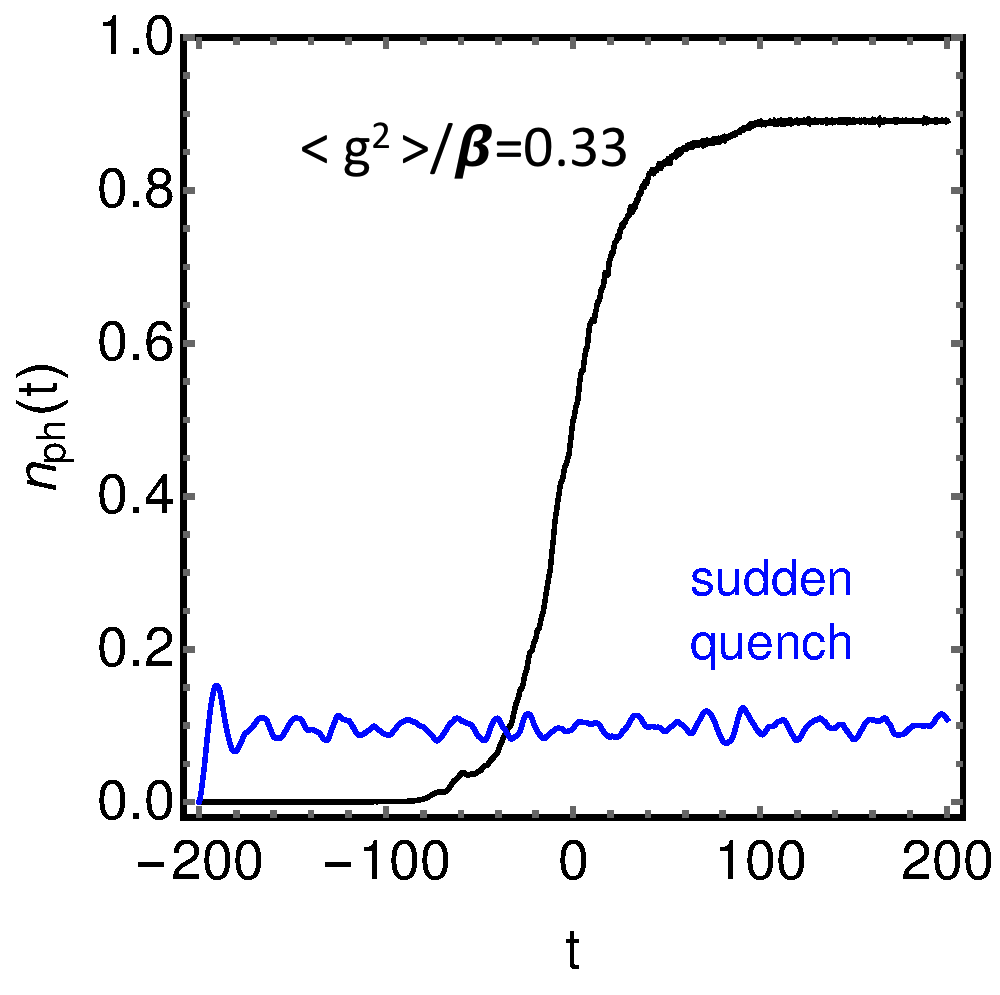}}
\caption{ Comparison of the normalized average number of photons $n_{ph}\equiv N_{ph}/N$ with $N_{ph}\equiv \la \hat a^\dag \hat a \ra$ emitted by $N=8$ initially up-polarized spins with the Hamiltonian (\ref{spectrumTC-fig}) after  a sudden quench (blue curve) that places $\omega=\la \varepsilon_k \ra$ at $t=-200$ and keeps $\omega$ constant afterwards;  and during a linear sweep of the frequency (black curve) according to Eq.~(\ref{linearT}) with $\beta=0.03$. 
Here, $\varepsilon_k$ and $g_k$ are random Gaussian numbers with  $\la \varepsilon_k \ra =0$, $\sqrt{{\rm var}(\varepsilon_k)}=1$,
$\la g_k \ra =0.1$, and  $\sqrt{ {\rm var} (g_k)}=0.02$. Both curves are averaged over 10 different realizations of random parameters.}
\label{nb-fig}
\end{figure}
%%%%%%%%%%%%%%%%%%%%%%%%%%%%%%%%%%%%%%%%%%%%%
Nevertheless, consider Fig.~\ref{nb-fig} that shows results of our numerical simulations for $N=8$ spins and parameters that mimic strong disorder \cite{note}. It demonstrates, that even if the strict adiabatic regime is not reached, the linear sweep of the frequency still produces many more photons than a sudden (Heaviside $\theta$-function) shift of the cavity mode frequency to the center of the resonance. Such a strong photon emission induced by a linearly time-dependent field is similar  to the one in the model without any inhomogeneous broadening \cite{altland3}. Hence, strong static disorder may not be  a problem for stimulated emission by the protocol (\ref{linearT}).

Here, we suggest that NV$^-$ centers inside a half-wavelength cavity can be used to store energy in spin excitations and release it on demand locally as a pulse of coherent radiation after a linear {\it nonadiabatic} frequency chirp (\ref{linearT}). We will show that if the photon loss rate is reasonably weak and $N$ is reasonably large, then the linear frequency chirp leads to collective radiation emission, with optical parameters being close to the perfect collective emission by all spins at no static disorder. This process, however, requires specific tuning of the sweeping rate $\beta$ because too fast changes are ignored by spins, whereas collective effects at slow sweeps are affected by  photon losses. Our theory will thus identify the condition for observation of this collective radiation effect. 

Our approach is based on the recently found solution of the model (\ref{dtcm1}) with a time-dependent frequency (\ref{linearT}) and constant couplings:  $g_k=g,\ \forall k$
\cite{DTCM,DTCM1,commute,yuzbahsyan}. The latter assumption is not essential because the main disorder problems originate from the distribution of $\varepsilon_k$, which keeps most of the two-level systems to be off-resonance at any given frequency of the cavity field. 
In \cite{DTCM1}, two of us derived the state-to-state transition probabilities for the sweep from large negative to large positive values, and derived the large-$N$ limit of this solution. However, further analysis was based on testing several predictions that were derived for $\varepsilon_k=0,\,\forall\,k$ in the large-$N$ limit by approximate methods \cite{altland3}. The latter work identified three regimes: with low and high emission of bosons, and an intermediate regime that could not be studied by well-justified methods. 

The exact solution, however, becomes particularly valuable when approximate methods of \cite{altland3} do not apply, i.e., at strong inhomogeneous broadening and moderate frequency sweeping rates. 
Let us now focus precisely on this poorly understood regime, where the model's solution will reveal a critical phenomenon.  According to \cite{DTCM1}, if an ensemble of $N$ spins is initially polarized ``up" and the optical mode is initially empty, then the probability of emitting $n$ photons after the linear chirp (\ref{linearT}) is given by the distribution:
\be
P_{n}=x^{N-n}(x^{N-n},x)_n, \quad x\equiv e^{-2\pi g^2/\beta}, 
\label{Pn}
\ee
where $(a,x)_k\equiv \prod_{j=0}^{k-1}(1-ax^j)$ is the q-Pochhammer symbol.

%%%%%%%%%%%%%%%%%%%%%%%%%%%%%%%%%%%%%%%%%%%%%
\begin{figure}[!htb]
\scalebox{0.4}[0.4]{\includegraphics{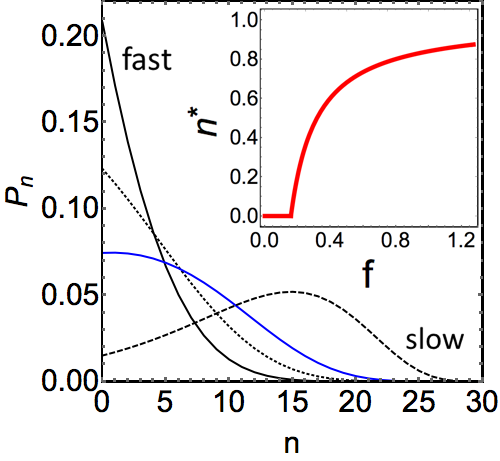}}
\caption{Distribution $P_n$ for $N=30$ spins and the ratio $\beta/g^2$ being  $4N$ (solid black), $3N$ (dotted black),  $2.42N$ (solid blue), and $1.5N$ (dashed black). Lines connect discrete points for better visibility. The blue curve corresponds to the critical $\beta$ at which the distribution maximum starts moving from $N^*=0$ to the right. The inset shows normalized position of the maximum, $n^*=N^*/N$, as a function of  $f=g^2N/(\beta \log_e N)$ in the limit $N\rar \infty$. Discontinuity of $n^*(f)$ marks the position of a phase transition.}
\label{dist-fig}
\end{figure}
%%%%%%%%%%%%%%%%%%%%%%%%%%%%%%%%%%%%%%%%%%%%%
This distribution for $N=30$ is shown in Fig.~\ref{dist-fig} for several different  $\beta$. 
A phase transition means discontinuous behavior of some measurable characteristic as a function of a control parameter. In our case, such a  transition can be defined even for a finite $N$. Namely, Fig.~\ref{dist-fig} shows that at fast sweeping rates (large $\beta$), the probability distribution $P_n$ has the maximum at $N^*=0$. However, at certain finite $\beta$, position of the maximum, $N^*$,  becomes not easy to define (blue curve), and at lower $\beta$ the maximum moves to a finite value $N^*>0$. Hence, position of the maximum of the distribution does not change continuously with $\beta$, which is a signature of a phase transition, in which the sweeping rate $\beta$ plays a role of the control parameter.  

One may wonder why this transition can be defined even for a finite $N$ because discontinuities in physical characteristics usually appear in the thermodynamic limit. The reason is that the probability distribution cannot be obtained by one measurement. To estimate the position of the maximum of $P_n$ definitely, one must  perform infinitely many measurements. Such phase transitions ``at fluctuation level" have attracted lot of interest recently because of advances of measurement techniques that probe fluctuations \cite{pt1,pt2,pt3}.  
For experiments, however, we do not propose to do many repetitions because, as we show below, this phase transition has much more accessible manifestations in the true thermodynamic limit $N\gg 1$. 

First, we note that the discontinuity in the position of the maximum dependence on $\beta$ survives even for a large $N$, when a continuous approximation to $P_n$ is applied. Let us define the normalized position of the maximum for the distribution (\ref{Pn}): $n^*=N^*/N$. 
According to \cite{DTCM1}, in the continuous limit, if $n^*>0$, then  
\be
n^*=1-\log_e(1-x)/(N\log_e x).
\label{max}
\ee
For large $N$, we further look for simplification of Eq.~(\ref{max}) by introducing an intensive  parameter 
\be
f\equiv g^2N/(\beta \log_eN),
\label{f-def}
\ee
which characterizes the inverse sweeping rate. Note that $f$ is dimensionless because the frequency sweeping rate $\beta \sim$[energy]$^2$. 
Substituting (\ref{f-def}) into (\ref{max}), we find for $N\gg 1$:
\be
n^*(f) =\frac{f-f_c}{f} + O(\log \log_e N/ \log_e N), 
\label{nf-curve}
\ee
where 
\be
 f_c=1/(2\pi)
\label{fc}
\ee
defines the critical point for the sweeping rate $\beta=\beta_c=2\pi g^2N/\log_e N$ at which the maximum of the probability distribution starts moving  from $n^*=0$. Thus, 
in the thermodynamic limit $N\rar \infty$, there are intensive characteristics $n^*$ and $f$, in terms of which the phase transition point (\ref{fc}) and the scaling (\ref{nf-curve})  do not depend on $N$. 
For  $\beta>\beta_c$,  $n^*=0$, and for $\beta<\beta_c$ we find that $n^*$ is growing and reaching values close to a unit value at $\beta \sim g^2N/\log_e N$, i.e. $N/\log_e N$ times larger than the value required to reach the adiabatic limit. The corresponding scaling near the critical point is 
\be
n^*(\beta) \sim  \Big\{  \begin{array}{l} (\beta_c-\beta)^\nu, \quad \beta<\beta_c, \quad \quad \nu =1,\\
 0, \quad \beta>\beta_c. \end{array}
 \label{powerlaw}
\ee

Although corrections in (\ref{nf-curve}) are suppressed only logarithmically, experiments can have, e.g., $N\sim 10^{12}$ NV-centers coupled to a microwave mode \cite{nv-strong}, for which $\log_e N \approx 27$, so the estimate (\ref{fc}) is close to the real value for such $N$. More precise position of 
the phase transition is given by equation $P_0=P_1$ with (\ref{Pn}), which leads to a nonlinear equation that defines the critical sweeping rate via $x_c\equiv e^{-2\pi g^2/\beta_c}$:
\be
x_c=1-x_c^N, 
\label{xc-def}
\ee
from which  the  finite size corrections to (\ref{fc})  can be developed.

\begin{figure}[!htb]
 \scalebox{0.334}[0.334]{\includegraphics{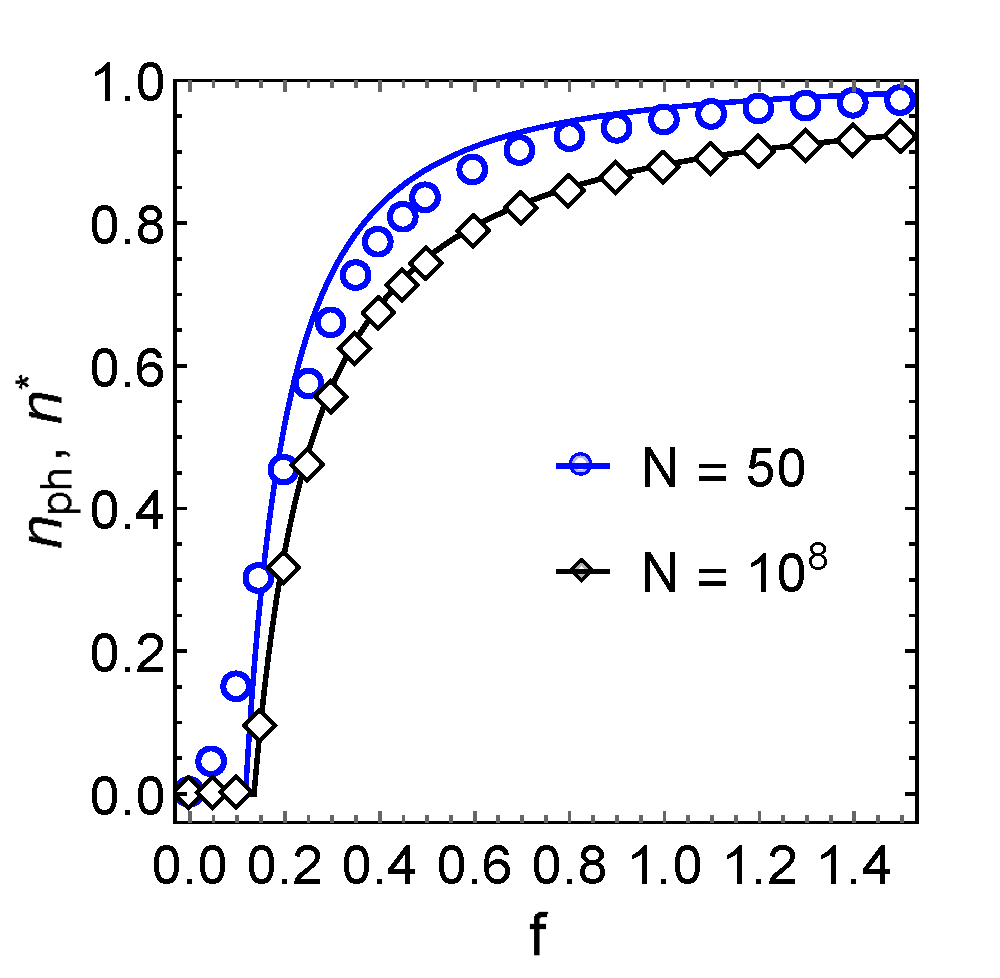}}
 \scalebox{0.334}[0.334]{\includegraphics{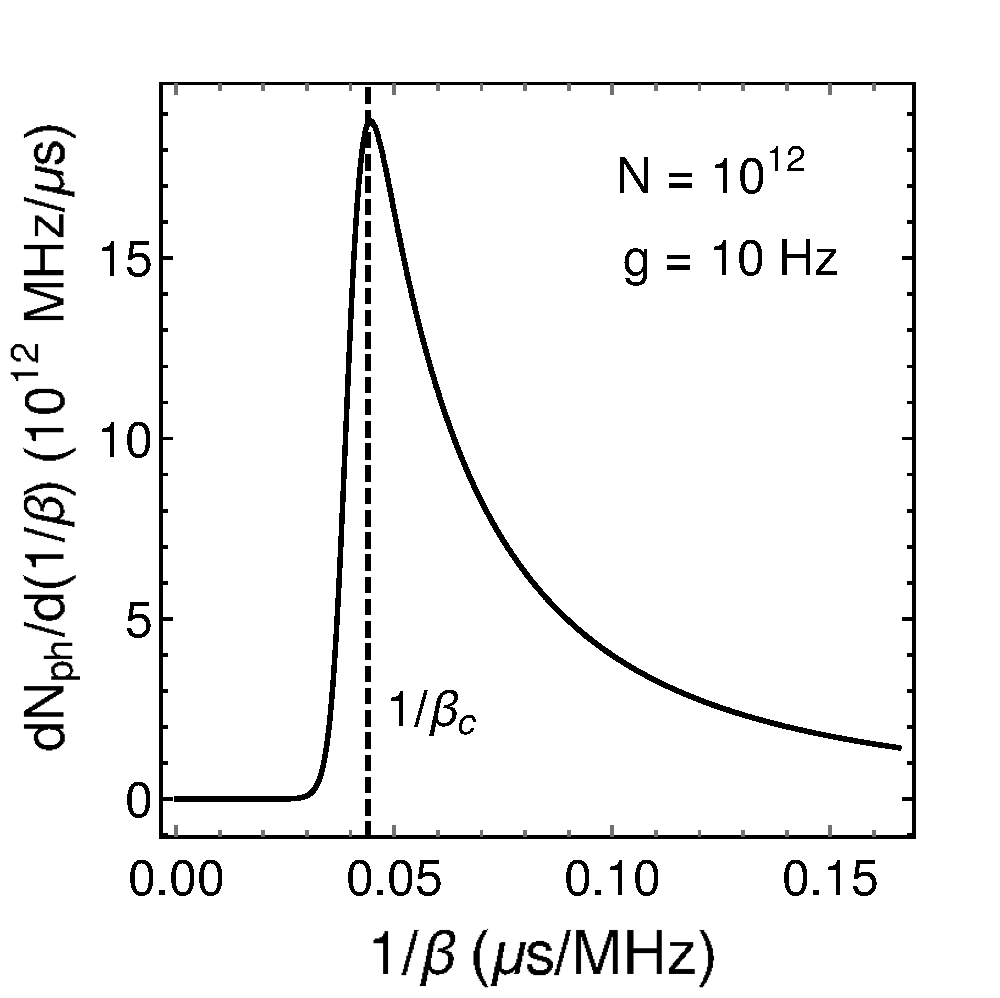}}
\caption{Left:  (discrete points) $n_{ph}(f)$ and   (solid curves) $n^*(f)$ given by~(\ref{max}), for  $N=50$ and $N=10^8$ and the distribution~(\ref{Pn}). 
The discontinuity  at $f=f_c$ emerges for $n_{ph}(f)$ as $N\rar \infty$.  Right: the derivative of the number of produced photons, $dN_{ph}/d(1/\beta)$, versus the inverse sweeping rate. The sharply rising piece of the curve  before the critical $1/\beta_c$  (from Eq.~(\ref{xc-def})) becomes vertical in the limit $N\rar \infty$. }
\label{avg-fig}
\end{figure}

Second, the mean number of the emitted photons, $N_{ph}=\la \hat a^\dag \hat a \ra$, is  more relevant  to experiment. 
For $N\gg 1$, this number should be close to the maximum of the peak value $N^*$. Therefore, in the thermodynamic limit ($N\rar \infty$), $N_{ph}$ should behave as the position of the maximum from the inset to Fig.~\ref{dist-fig}: $N_{ph}\approx N^*$.  To check this hypothesis, we calculated the 
 normalized average number of the emitted photons, $n_{ph}\equiv N_{ph} /N$ for the distribution (\ref{Pn}) numerically. Its dependence on $f$ for  different $N$ is shown in Fig.~\ref{avg-fig}(Left), where we also compare $n_{ph}(f)$ with positions of the maximum $n^*(f)$ at the same $N$.  At  $N=50$, the difference between $n_{ph}(f)$ and $n^*(f)$ is noticeable but 
at $N =10^8$, the average number of the emitted photons follows predictions of Eq.~(\ref{max}), which also do not change much for higher $N$. 

Thus, $n_{ph}(f)$ develops a singularity near $f=f_c$ with the same power-law behavior as in (\ref{powerlaw}). Moreover, away from the critical point in the ``slow" phase, it  reaches the values  $n_{ph}\approx1$. That is,  at much faster than adiabatic sweeping rates, the majority of two-level systems emit photons. So, the behavior of the system in this phase is similar to the adiabatic one despite the adiabatic conditions are not even close to be satisfied.
The latter behavior was noticed in \cite{altland3} using semiclassical methods but only for the  case without disorder (${\rm var}(\varepsilon_k)=0$), for which the collective emission would be expected even  at fixed resonant  $\omega$.

 Figure.~\ref{avg-fig}(Right) shows the critical behavior that we propose to observe  experimentally in two-level systems coupled to a cavity mode with a tunable frequency. The experimentally controlled parameter 
here is the sweeping rate $\beta$. In order to reach the critical point, this parameter should be
\be
\beta_c =g^2N/(f_c \log_eN) = 2\pi g^2N/\log_eN. 
\label{bc1}
\ee
%%%%%%%%%%%%%%%%%%%%%%%%%%%%%%%%%%%%%%%%%%%%%
%%%%%%%%%%%%%%%%%%%%%%%%%%%%%%%%%%%%%%%%%%%%%
The characteristic time of passage through the range of energies with resonances is given by $2T=2\sqrt{{\rm var}(\varepsilon_k)}/\beta_c$.  This should be smaller than the photon lifetime in the cavity in order not to destroy the collective emission effect.   The photon loss rate is $\bar{\varepsilon}_k/Q$, so, the phase transition is observable when $2\bar{\varepsilon}_kT/Q<1$, or
\be
\bar{\varepsilon}_k \sqrt{{\rm var}(\varepsilon_k)} \log_e N/(\pi Q g^2 N) <1.
\label{cond1}
\ee

For NV$^-$ centers, we find then from (\ref{cond1}) that the minimal number of 
defects inside the cavity with $Q=10^6$ should be $N=N_{\rm min}\sim 6\cdot 10^8$. At such parameters, $g\sqrt{N} \approx 0.3$MHz, and $\beta_c \sim 10^{4}$MHz/sec. That is, although the system is in the weak coupling regime, the collective radiation for the majority of initially excited spins can be induced by a frequency chirp that crosses the region with resonances during time $ 2\sqrt{{\rm var}(\varepsilon_k)}/\beta_c  \sim 10^{-4}$sec and produces $\sim10^6-10^8$ photons that  leave the cavity during  $\tau_c\sim 0.3\cdot 10^{-3}$sec. 
This time estimate is below the quantum spin coherence time $T_2\sim 1$ms, which is  found for NV-centers in diamond at various temperatures and sample quality conditions  \cite{long-coherence-NV1,long-coherence-NV2}. 

Our phase transition can be understood semiclassically. Namely, if one makes the frequency chirp sufficiently slow to flip just $\sim {\rm log}_eN$ spins, then the positive feedback creates an avalanche of photon emissions that propagates through the rest of the spins.  
This mechanism does not rely on massive entanglement or other quantum correlations that could be sensitive to decoherence. Moreover, Landau-Zener transitions at fast sweeping rates are particularly robust against decoherence \cite{prokofev}.  
Hence, even for fast decoherence $T_2\sim 1\mu$s, which is often reported at room temperature, we expect that our phase transition will be found. %Our theoretical methods  do not apply to  such small $T_2$, so the quantitative details for spin coherent and incoherent regimes may be different and remain to be understood.

Thus, we conclude that demonstration of a coherent light pulse that emerges below a critical frequency sweeping rate, $\beta<\beta_c$, in the weak coupling regime, $g\sqrt{N}<\sqrt{{\rm var}(\varepsilon_k)}$, is well within reach at cryogenic temperatures and is likely observable even at room temperature.  Our calculations also apply to the strong coupling regime  but collective emission at strong coupling can be demonstrated by standard means \cite{nv-strong}.

There are  potential applications of this effect in  circuit-QED. Although a maser on NV centers already exists \cite{nv-maser1}, collective radiation during the frequency chirp (\ref{linearT}) can be induced by a relatively small number of emitters. It should be valuable to create such local coherent radiation pulses  that can exchange information between  elements of a quantum circuit. %Our estimates show that, indeed, it is possible  to release energy of  a mesoscopic number of  spin-polarized NV$^-$ centers  as a high intensity pulse of coherent radiation even at strong inhomogeneous broadening and weak coupling.  
 
The dynamic phase transition that we introduced is also interesting from the point of view of general physics. It is a different phenomenon from commonly known  dynamic phase transitions  \cite{dyn-pt1,dyn-pt2,dyn-pt-exp}, in which sharp changes are encountered at some evolution time moment. In our case, the word ``dynamic" is justified rather by the fact that the control parameter $\beta$ characterizes  the deviation of the system from adiabaticity. 

Discontinuous behavior of the order parameter and its scaling near the critical point make the ``fast-slow" transition  similar to standard second order phase transitions  in equilibrium systems. However, we point to one  difference. Namely,  phase transitions are usually associated with spontaneous symmetry breaking during minimization of some function of parameters, such as  the free energy, dissipation rate, or a Hamiltonian ground energy. Curiously, within the driven Tavis-Cummings model the critical behavior is found rather as a result of purely coherent evolution.
As in the theory of standard phase transitions, it is expected that this behavior is not unique for our system but is a signature of a universality class. There can  be  similar phase transitions  with a different scaling near the critical point, which we  hope  will be found too.

\acknowledgments
We thank A. Bleszynski Jayich and A. Saxena for useful discussions. 
This work was supported by the U.S. Department of Energy, Office of Science, Basic Energy
Sciences, Materials Sciences and Engineering Division, Condensed Matter Theory Program (V.Y. C. and N.A.S.), and by the J. Michael Kosterlitz Postdoctoral Fellowship at Brown University (C.S.).
N.A.S. and A.P. also thank the support from the LDRD program at LANL.

\end{document}